\documentclass[12pt,]{article}
 \usepackage{graphicx}
 \usepackage[cp1251]{inputenc}
 \usepackage{indentfirst}

\begin{document}

{\small\bf\noindent ISSN 1063-7737, Astronomy Letters, 2014, Vol.
40,
No. 6, pp. 352–359. \copyright Pleiades Publishing, Inc., 2014.\\
Original Russian Text \copyright V.V. Bobylev, A.T. Bajkova, 2014,
published in Pis'ma v Astronomicheskii Zhurnal, 2014, Vol. 40, No.
6, pp. 396-403.}

{\noindent\_\_\_\_\_\_\_\_\_\_\_\_\_\_\_\_\_\_\_\_\_\_\_\_\_\_\_\_\_\_\_\_\_\_\_\_\_\_\_\_\_\_\_\_\_\_\_\_\_\_\_\_\_\_\_\_\_\_\_\_\_\_\_\_\_\_\_\_\_\_\_\_\_\_\_\_\_\_\_\_\_\_\_\_\_\_\_\_\_\_\_\_\_\_\_\_\_\_\_\_\_\_\_\_\_\_\_\_\_\_}

\bigskip

 \centerline {\Large\bf Search for Kinematic Siblings of the Sun Based on }
 \centerline {\Large\bf Data from the XHIP Catalog}

 \bigskip
 \centerline {V. V. Bobylev$^{1, 2}$ and A. T. Bajkova$^1$}

 \bigskip
 \centerline {\small $^{1}$\it Pulkovo Astronomical Observatory, Russian Academy of
 Sciences,}
 \centerline {\small \it Pulkovskoe sh. 65, St. Petersburg, 196140 Russia}

 \centerline {\small $^{2}$\it Sobolev Astronomical Institute, St. Petersburg State University,}
 \centerline {\small \it Universitetskii pr. 28, Petrodvorets, 198504 Russia}

 \bigskip
\centerline { Received November 11, 2013}

 \bigskip

 {\noindent\small
 {\bf Abstract} — From the XHIP catalogue, we have selected 1872 F–G–K stars with relative parallax measurement
errors $<20$\% and absolute values of their space velocities
relative to the Sun $<15$ km s$^{-1}$. For all these stars, we
have constructed their Galactic orbits for 4.5 Gyr into the past
using an axisymmetric Galactic potential model with allowance made
for the perturbations from the spiral density wave. Parameters of
the encounter with the solar orbit have been calculated for each
orbit. We have detected three new stars whose Galactic orbits were
close to the solar one during a long time interval in the past.
These stars are HIP 43852, HIP 104047, and HIP 112158. The
spectroscopic binary HIP 112158 is poorly suited for the role of a
kinematic sibling of the Sun by its age and spectroscopic
characteristics. For the single star HIP 43852 and the multiple
system HIP 104047, this role is quite possible. We have also
confirmed the status of our previously found candidates for close
encounters, HIP 47399 and HIP 87382. The star HIP 87382 with a
chemical composition very close to the solar one is currently the
most likely candidate, because it persistently shows close
encounters with the Sun on time scales of more than 3 Gyr when
using various Galactic potential models both without and with
allowance made for the influence of the spiral density wave. }

\bigskip\noindent DOI: 10.1134/S1063773714060012


\newpage
\section*{INTRODUCTION}
Searching for stars that could be formed with the Sun in a common
cluster is of great interest for studying the long-term evolution
of the Solar system in the Galaxy. This problem has been solved in
recent years based on stellar proper motions (Brown et al. 2010)
and total stellar space velocities (Bobylev et al. 2011).

The stars born together are expected to retain a similar chemical
composition for a long time (Bland- Hawthorn and Freeman 2004;
Bland-Hawthorn et al. 2010). This provides a basis for the method
of searching for chemical counterparts by analyzing spectroscopic
data to determine the abundances of various chemical elements.
Thus, when searching for kinematic siblings of the Sun, we proceed
from the fact that their chemical composition should be maximally
close to the solar one.

Not much is known about such specific characteristics of the
protosolar cluster as its mass, density, initial size, and the
number of members. Therefore, the paper by Pfalzner (2013), where
the possible developmental paths of such a cluster are discussed
and extensive literature on this question is given, is of great
interest. The author argues that the Sun most likely formed in a
cluster like an OB association with an initial number of members
of at least 1000. Such clusters are known to remain
gravitationally bound structures for a very short time (several
Myr) and to dissipate rather quickly under the influence of the
Galactic gravitational field. However, the density of such
structures is not high enough for the protoplanetary disks to be
disrupted through mutual encounters of stars.

According to the estimates by Portegies Zwart (2009), 10-60 stars
from the protosolar star cluster that initially contained $\sim
10^3$ members can be currently contained in the solar neighborhood
with a radius of 100 pc. Mishurov and Acharova (2011) showed that
the influence of the spiral density wave could lead to a
significant scattering of members of an initially compact cluster.
According to their estimate, for about a hundred stars to be
observable in the solar neighborhood with a radius of 100 pc, the
protosolar cluster must contain at least 104 members. Thus, it is
very important to take into account the influence of the spiral
density wave in our problem.

This study is aimed at searching for potential candidates that
could be formed with the Sun in a common cluster by analyzing
their 3D kinematics. For this purpose, we search for stars whose
Galactic orbits remained close to the solar one during a long time
interval in the past. Possibilities for a new search arise from
the appearance of the XHIP compilation (Anderson and Francis
2012), which allows the space velocities of $\sim 46 000$
Hipparcos stars to be analyzed.

\section*{DATA}
In the XHIP catalogue, the parallaxes were taken from the revised
version of the Hipparcos catalogue (van Leeuwen 2007) and the
stellar proper motions were taken from the Hipparcos and Tycho-2
(Hog et al. 2000) catalogues or their combination. The radial
velocity measurements are available for 46 392 stars. Previously
(Bobylev et al. 2011), we worked with the catalogue by Gontcharov
(2006) catalogue, which contains the radial velocities for 35 493
Hipparcos stars. There are more stars with measured radial
velocities in the XHIP catalogue; new measurements were included
for a number of single stars and, what is very important, the data
for spectroscopic binaries were checked against the updatable SB9
bibliographic database (Pourbaix et al. 2004).

We selected 1872 F, G, and K stars with relative parallax
measurement errors $\sigma_\pi/\pi<20\%$ and absolute values of
their total space velocities relative to the Sun $\sqrt
{U^2+V^2+W^2}<15$~ km s$^{-1}$. Compared to our previous paper
(Bobylev et al. 2011), we relaxed significantly the selection
criteria and, therefore, the number of stars for our analysis
increased by an order of magnitude.

For all these stars, we constructed their Galactic orbits for 4.5
Gyr into the past using the three component axisymmetric Galactic
potential model from Fellhauer et al. (2006) by taking into
account the perturbations from the spiral density wave (Fernandez
et al. 2008). For each stellar orbit, we calculated such
parameters of its encounter with the solar orbit as the relative
distance $d$ and the relative velocity $dV$.

It can be assumed that the initial size of the protosolar cluster
could be about 10–20 pc. Given that there are errors in the
original observational data, we consider the falling of a star
into the zone $d<50$ pc in the time interval $t<-3$ Gyr at a
relative velocity $dV$ of several km s$^{-1}$ an encounter.

\section*{ORBIT CONSTRUCTION}
\subsection*{Potential Model}
We calculated the Galactic orbits of the Sun and stars by solving
the following system of equations of motion based on a realistic
Galactic gravitational potential model:

\begin{equation}
\ddot{\xi}=-\frac{\partial\Phi}{\partial\xi}-\Omega^2_0(R_0-\xi)-2\Omega_0\dot{\eta},
\end{equation}
 $$
\ddot{\eta}=-\frac{\partial\Phi}{\partial\eta}+\Omega^2_0\eta+2\Omega_0\dot{\xi},
 $$$$
\ddot{\zeta}=-\frac{\partial\Phi}{\partial\zeta},
 $$
where $\Phi$ is the Galactic gravitational potential; the
$(\xi,\eta,\zeta)$ coordinate system centered on the Sun rotates
around the Galactic center with a constant angular velocity
$\Omega_0$, with the $\xi$ axis being directed toward the Galactic
center, the $\eta$ axis being in the direction of Galactic
rotation, and the $\zeta$ axis being directed toward the North
Galactic Pole; $R_0$ is the Galactocentric distance of the Sun.

Bobylev et al. (2011) searched for candidates using the Galactic
potential model by Allen and Santillan (1991) rigidly tied to
$R_0=8.5$ kpc. According to the review, for example, by Foster and
Cooper (2010), the present-day value of this distance is
$R_0=8.0\pm0.4$ kpc. Therefore, in this paper, we used a potential
model that approximated more closely the currently available data.
This is the model by Fellhauer et al. (2006):

 \begin{equation}
 \Phi=\Phi_{halo}+\Phi_{disk}+\Phi_{bulge},
 \label{model}
 \end{equation}
where
\begin{itemize}
 \item the halo is represented by a potential dependent
on the cylindrical Galactic coordinates $R$ and $Z$ as
$\Phi_{halo}(R,Z)=\nu_0^2\ln(R^2+d^2)$, where $\nu_0=134$ km
s$^{-1}$ and $d = 12$ kpc;
 \item the disk is represented by the Miyamoto–
Nagai (1975) potential dependent on the same coordinates:
$\Phi_{disk}(R,Z)=-GM_d(R^2+(b+(Z^2+c^2)^{1/2})^2)^{-1/2}$, where
the disk mass $M_{d}=9.3\times10^{10}$ $M_{\odot}$, $b=6.5$ kpc,
and $c=0.26$ kpc;
 \item the bulge is represented by the Hernquist (1990)
potential, $\Phi_{bulge}(R)=-GM_{b}/(R+a)$, where the bulge mass
$M_{b}=3.4\times10^{10}$ $M_{\odot}$ and  $a=0.7$ kpc.
\end{itemize}
The Galaxy’s circular rotation velocity at the distance $R_0=8$
kpc is 220 km s$^{-1}$. The components of the Sun’s peculiar
velocity relative to the local standard of rest were taken to be
$(U_\odot,V_\odot,W_\odot)_{LSR}=(10,11,7)$ km s$^{-1}$ according
to the results by Bobylev and Bajkova (2010), in agreement with
the results by Sch\"{o}nrich et al. (2010). We take into account
the Sun’s displacement from the Galactic plane $Z_\odot=17$ pc
(Joshi 2007).

In the case where the spiral density wave is taken into account
(Lin and Shu 1964; Lin et al. 1969), the following term (Fernandez
et al. 2008) is added to the right-hand side of Eq. (2):

\begin{equation}
 \Phi_{sp} (R,\theta,t)= A\cos[m(\Omega_p t-\theta)+\chi(R)],
\end{equation}
where
 $$
 A= \frac{(R_0\Omega_0)^2 f_{r0} \tan i}{m},
 $$$$
 \chi(R)=- \frac{m}{\tan i}
 \ln\biggl(\frac{R}{R_0}\biggr)+\chi_\odot.
 $$
Here, $A$ is the amplitude of the spiral density wave potential;
$f_{r0}$ is the ratio of the radial perturbation component from
the spiral arms to the Galaxy's total attraction; $\Omega_p$  is
the wave pattern speed; $m$ is the number of spiral arms; $i$ is
the pitch angle of the arms, $i<0$ for a winding pattern; $\chi$
is the phase of the radial wave, the arm center then corresponds
to the phase $\chi=0^\circ$; and $\chi_\odot$ is the Sun's radial
phase in the spiral density wave.

\subsection*{Parameters of the Spiral Density Wave}
The known parameters of the spiral density wave are very
unreliable. As analysis of the spatial distribution of young
Galactic objects (OB stars, star-forming regions, young open
clusters, or HI clouds) shows, a two-armed, three-armed, and
four-armed pattern is possible. More complex models are also known
(Vall\'{e}e 2008). This ambiguity is related to large errors in
the distances (photometric and kinematic) to distant tracers of
the spiral pattern.

Bobylev and Bajkova (2014, 2013b) used data on 80 Galactic masers
with known trigonometric parallaxes measured by VLBI with an
error, on average, of less than 10\% to estimate the pitch angle
of the Galactic spiral arms. These masers are associated with very
young objects located in regions of active star formation. They
are distributed in a wide range of distances, up to $R=20$ kpc. In
the outer arm, we invoked data on 12 very young star clusters with
their distances estimated from infrared photometry. Based on a
direct estimation method, we showed that the model of a four-armed
spiral pattern ($m=4$) with the pitch angle
$i=-13^\circ\pm1^\circ$ is most likely realized in the Galaxy.

The Sun’s radial phase in the spiral density wave depends on the
age of sample stars. For example, based on a sample of distant
massive OB stars, we found $\chi_\odot=-90^\circ$ (Bobylev and
Bajkova 2013c); from masers, we obtained the phase from
$\chi_\odot=-140^\circ$ (Bobylev and Bajkova 2014) to
$\chi_\odot=-160^\circ$ (Bobylev and Bajkova 2013a); for Cepheids,
this parameter exhibits a noticeable age trend, from
$\chi_\odot\approx-150^\circ$  for the youngest ones to
$\chi_\odot\approx-240^\circ$ for the oldest ones (Bobylev and
Bajkova 2012).

The estimates of the spiral pattern speed $\Omega_p$ lie within
the range 15–30 km s$^{-1}$ kpc$^{-1}$ (Popova and Loktin 2005;
Gerhard 2011; Bobylev and Bajkova 2012). We chose a moderate
value, $\Omega_p=20$ km s$^{-1}$ kpc$^{-1}$, for our model
problem. Since the angular velocity of Galactic rotation in our
potential (2) is $\Omega_0=V_0/R_0=27.5$ km s$^{-1}$ kpc$^{-1}$,
the Sun lies within the corotation radius, where the influence of
the spiral density wave is strongest (Mishurov and Acharova 2011).

According to the classical approach in the linear theory of
density waves (Yuan 1969), the ratio $f_{r0}$ lies within the
range 0.04–0.07 and its most probable value is $f_{r0}=0.05$
(Fernandez et al. 2008). The upper limit $f_{r0}=0.07$ is
determined by the dispersion of young objects observed in the
Galaxy (the dispersion at $f_{r0}=0.07$ should reach 25 km
s$^{-1}$, which exceeds a typical observed value of 10–15 km
s$^{-1}$). Note that Mishurov and Acharova (2011) took a fairly
large value, $f_{r0}=0.1,$ to demonstrate the influence of the
spiral density wave on the motion of protosolar cluster members;
the Sun was located on the corotation circle and, as a result,
they estimated the maximum effect.

All of the aforesaid gives us reason to adopt the following
parameters of the spiral density wave:
 \begin{equation}
 \begin{array}{lll}
 m=4,\\
 i=-13^\circ,\\
 f_{r0}=0.05,\\
 \chi_\odot=-140^\circ,\\
 \Omega_p=20~\hbox {km s$^{-1}$ kpc$^{-1}$}
 \label{param-spiral}
 \end{array}
 \end{equation}
as the initial approximation. Some of them, in particular,
$\chi_\odot$ and $\Omega_p$, can be varied if necessary.

\begin{figure}[t]{
\begin{center}
 \includegraphics[width=165mm]{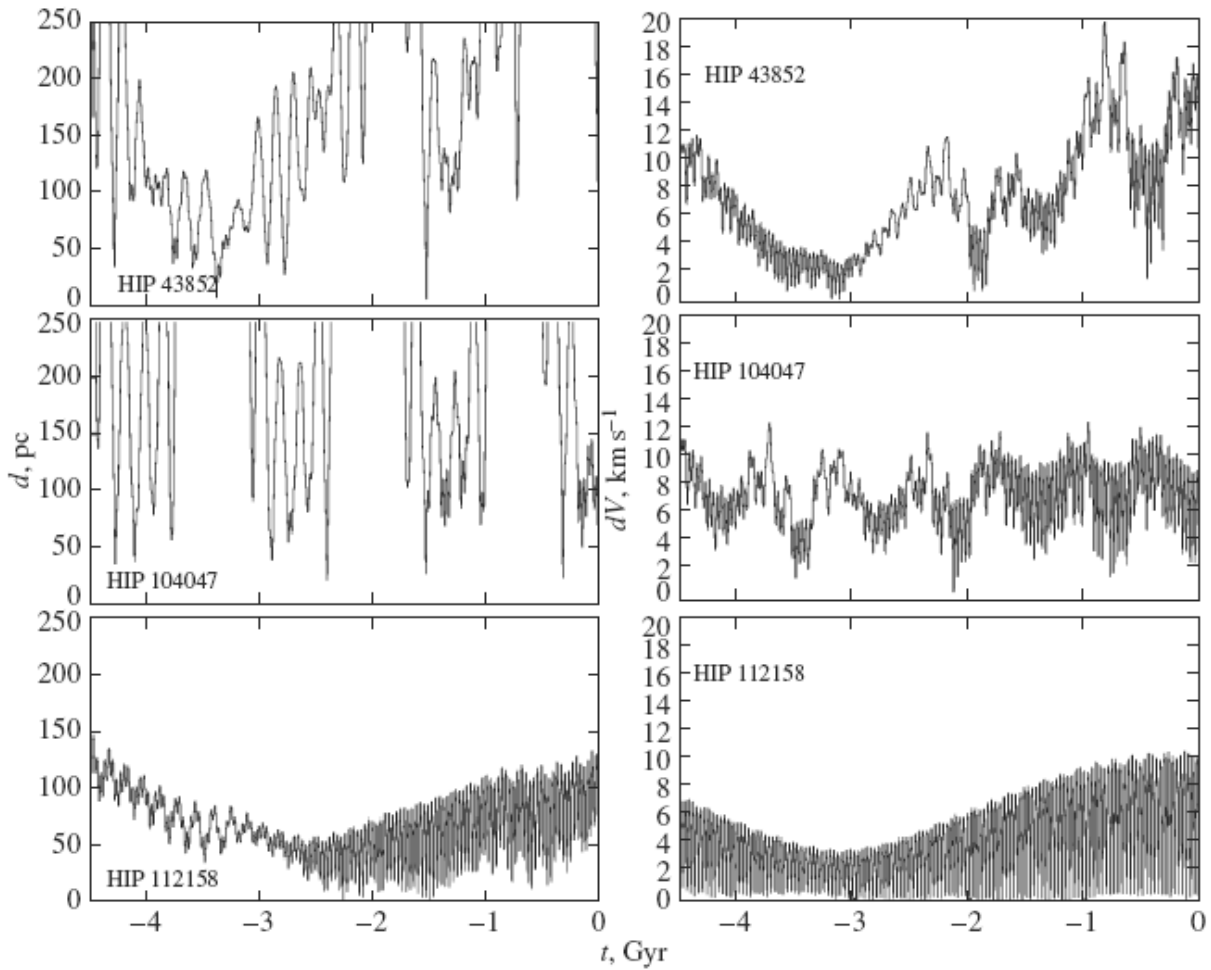}
 \caption{Parameters of the encounter $d$ and  $dV$ of the three stars found with the solar orbit obtained by taking into account the
influence of the spiral pattern with parameters (4) versus time.
 }
\label{f1}
 \end{center}}
  \end{figure}

\section*{RESULTS AND DISCUSSION}
\subsection*{The Model of a Four-Armed Spiral Pattern}
Among the selected 1872 F, G, and K stars, only five stars are of
interest: HIP 43852, HIP 104047, HIP 112158, HIP 47399, and HIP
87382. We failed to find close encounters with the solar orbit for
all the remaining stars.

The first three stars are completely new candidates for the Sun’s
kinematic siblings. The parameters of their encounter with the
solar orbit obtained by taking into account the influence of the
four-armed spiral pattern are given in Fig. ~\ref{f1}. The
physical characteristics of these stars are presented in the table
\ref{Tab-fiz}. Their analysis leads us to conclude that HIP~112158
is poorly suited for the role of the Sun’s kinematic sibling by
its age, while this role is quite possible for HIP~43852 and
HIP~104047.

It is interesting to check the status of our previously detected
candidates for close encounters, HIP~47399 and HIP~87382, with
partially new initial data in a new potential with parameters
~(\ref{param-spiral}). For these stars, there are the
corresponding plots of the encounter parameters constructed for a
four-armed pattern with $i=-10^\circ$, $f_{r0}=0.05$, and
$\chi_\odot=-117^\circ$ in Bobylev et al. (2011).

\begin{figure}[t]{
\begin{center}
 \includegraphics[width=165mm]{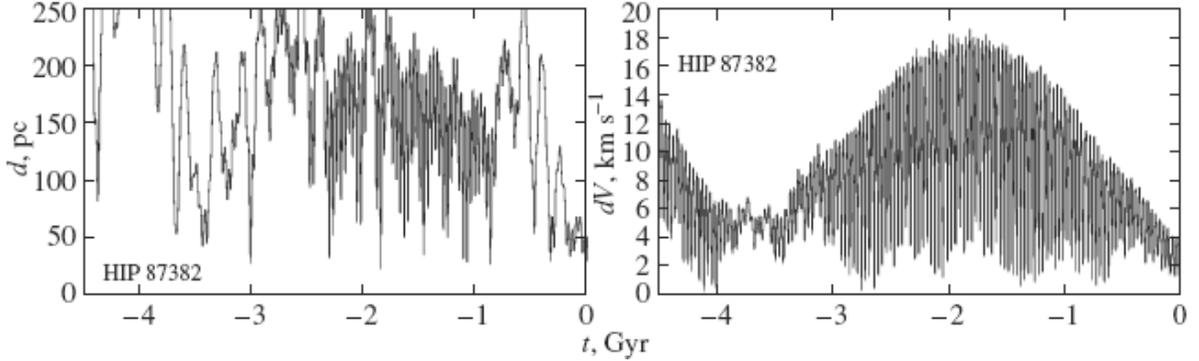}
 \caption{Parameters of the encounter $d$ and $dV$ of HIP~87382 with the solar orbit obtained by taking into account the influence
of the four-armed spiral pattern with parameters (4) versus time.
 }
\label{f2}
 \end{center}}
  \end{figure}

Note that the proper motions and radial velocities of these stars
in the XHIP catalogue (Anderson and Francis 2012) differ from
those that we used previously. The difference for HIP~87382 is not
large; its encounter parameters $d$ and $dV$ shown in
Fig.~\ref{f2}, differ only slightly from those derived previously
(Fig.~7 in Bobylev et al. 2011).

The difference in initial proper motions (in fact, between the
Hipparcos and Tycho-2 data) for HIP~47399 is significant, $\Delta
\mu_\alpha\cos\delta\approx3$ mas yr$^{-1}$ and $\Delta
\mu_\delta\approx0.5$ mas yr$^{-1}$. At the distance to the star
$r=72$ pc, this difference gives a change in the linear velocity
by $\Delta V=4.74 r \Delta \mu\approx1$ km s$^{-1}$, which affects
noticeably its orbital parameters; the statistics of encounters
with the Sun becomes poorer — there are only rare falls into the
neighborhood $d<150$ pc. Note that using the proper motions from
the Tycho-2 catalogue (less accurate than those in the Hipparcos
catalogue in terms of random errors) is justified when analyzing
the kinematics of double and multiple systems. Below, we provide
some of the physical characteristics for the candidates found.

\begin{table}
\caption{Physical characteristics of the stars}
  \label{Tab-fiz}
\begin{center}
{\small
\begin{tabular}{|r|c|r|c|c|c|c|c|c|}\hline
    HIP & Distance, pc & n    & Spectral type    & Age, Gyr  & [Fe/H]  & Mass,$M/M_\odot$ & Reference \\\hline
        43852 & 26          & 1     & G5V           & 6.6       & $-0.05$ & 0.83        & (1,2) \\
 104047 & 82          & 3     & G0V           & 4.6       & $+0.08$ & 1.1$+$0.74$+$0.71   & (2,6) \\
 112158 & 66          & 2     & G2II--III, SB &           & $-0.13$ &             & (3)   \\
        &             &       & G7II$+$A2(--) & 0.2       &         & 3.6$+$2.0   & (4,5) \\\hline

\end{tabular}
}
\end{center}
{\small
 $n$ is the number of components in the system. (1) Valenti and Fischer (2005); (2) Casagrande et al. (2011); (3) XHIP (Anderson and
Francis 2012); (4) Parsons (2004); (5) Massarotti et al. (2008);
(6) Tokovinin (2008).
 }
\end{table}

{\bf HIP~43852.} This is a fairly close single star with unusual
properties. For example, based on its kinematic properties, Montes
et al. (2001) attributed it to the fairly young (300 Myr) Ursa
Major moving group. However, despite its kinematic proximity to
the group, Nakajima and Morino (2012) attributed it to field stars
due to the absence of brightness variability as well as the
ultraviolet and X-ray radiation typical of young stars.

The age estimates for this star obtained by Casagrande et al.
(2011) by fitting to various (Padova and BASTI) isochrones turned
out to be contradictory, from $\sim 1$ to 13 Gyr.

The age estimate in the table was obtained by Valenti and Fischer
(2005) using the set of Yonsei-Yale isochrones (Demarque et al.
2004). The lower and upper age limits were 1.4 and 11.7 Gyr,
respectively, with the mean value being 6.6 Gyr. These authors
analyzed high-resolution ($R \approx 70000$) spectra taken with
the 10-m Keck telescope, the 4-m Anglo-Australian telescope of the
Siding Spring Observatory, and the 4-m telescope of the Lick
Observatory. Several fundamental parameters determined by these
authors include the effective temperature $T_{eff}=5465$ K; the
surface gravity log$(g)=4.72$ (where $g$ is in cm s$^{-2}$); the
rotational velocity $v$sin$i=3.7$ km s$^{-1}$; the relative
abundances of the following chemical elements: [M/H] = 0.02,
[Na/H] = 0.11, [Si/H] = 0.01, [Ti/H] = 0.09, [Fe/H] = 0.07, and
[Ni/H] = 0.04; the mean error in these indices is about 0.03. As
we see, the chemical composition of this star is fairly close to
the solar one.

{\bf HIP~47399.} This is a single star. When using the Hipparcos
proper motions, its encounters with the Sun are very close (Fig. 6
in Bobylev et al. 2011). Batista and Fernandes (2012) confirmed
its solar counterpart status from the viewpoint of chemical
composition. For this purpose, they analyzed all of the available
published data. All of this leads to the conclusion that the
status of this star as a candidate for the Sun’s kinematic
siblings is retained.

{\bf HIP~87382.} Valenti and Fischer (2005) determined a number of
fundamental parameters for this star: $T_{eff}=6162$ K,
log$(g)=4.29,$ $v$sin$i=3.6$ km s$^{-1}$, [M/H] = 0.02, [Na/H] =
0.02, [Si/H] =0.03, [Ti/H] = 0.06, [Fe/H] = 0.03, and [Ni/H] =
0.00. This shows that the chemical composition of HIP 87382
differs only slightly from the solar one, making this star the
most likely candidate for the Sun’s siblings. Therefore, we
carried out a detailed study of its encounters with the Sun using
various Galactic potential models (Fellhauer et al. 2006; Allen
and Santillan 1991; Bobylev and Bajkova 2014) both without and
with allowance made for the influence of the spiral density wave.
The parameters of the spiral density wave $i$ and $\chi_\odot$
varied within their error limits. We found fairly stable
encounters on time scales of more than 3 Gyr into the past.

{\bf HIP~104047.} This is the multiple (A + BC) system WDS
21047+0332 (Mason et al. 2001). The apparent separation between
components A and BC is $a=3.''18$ ($\approx$260 AU). The lower age
limit for the primary component and the metallicity have been
determined by various authors quite reliably (Nordstr\"{o}m et al.
2004; Casagrande et al. 2011). Thus, this system may be considered
a suitable candidate for the Sun's siblings by its kinematics and
chemical composition.

{\bf HIP~112158.}The star is a well-known ($\eta$ Peg, Matar)
spectroscopic binary (SB1) with a period of 817.5 days (Massarotti
et al. 2008). The apparent separation between the components is
$a=0.''045$ ($\approx$3 AU); the hot component may be a binary
(Parsons 2004). The dynamical mass estimate for the system,
$5.17\pm0.72~M_\odot$ (Malkov et al. 2012), agrees satisfactorily
with the estimates obtained by Parsons (2004) by fitting to
theoretical evolutionary tracks (see the table). For this system,
we used a very reliable value of the systemic radial velocity,
$V_\gamma=4.17\pm0.35$ km s$^{-1}$ (Massarotti et al. 2008).

Based on its kinematics and spectroscopic data, this system can be
attributed to the Sun's chance traveling companions, a traveling
companion for a short time interval, no more than 500 Myr in the
past.

\begin{figure}[t]{
\begin{center}
 \includegraphics[width=165mm]{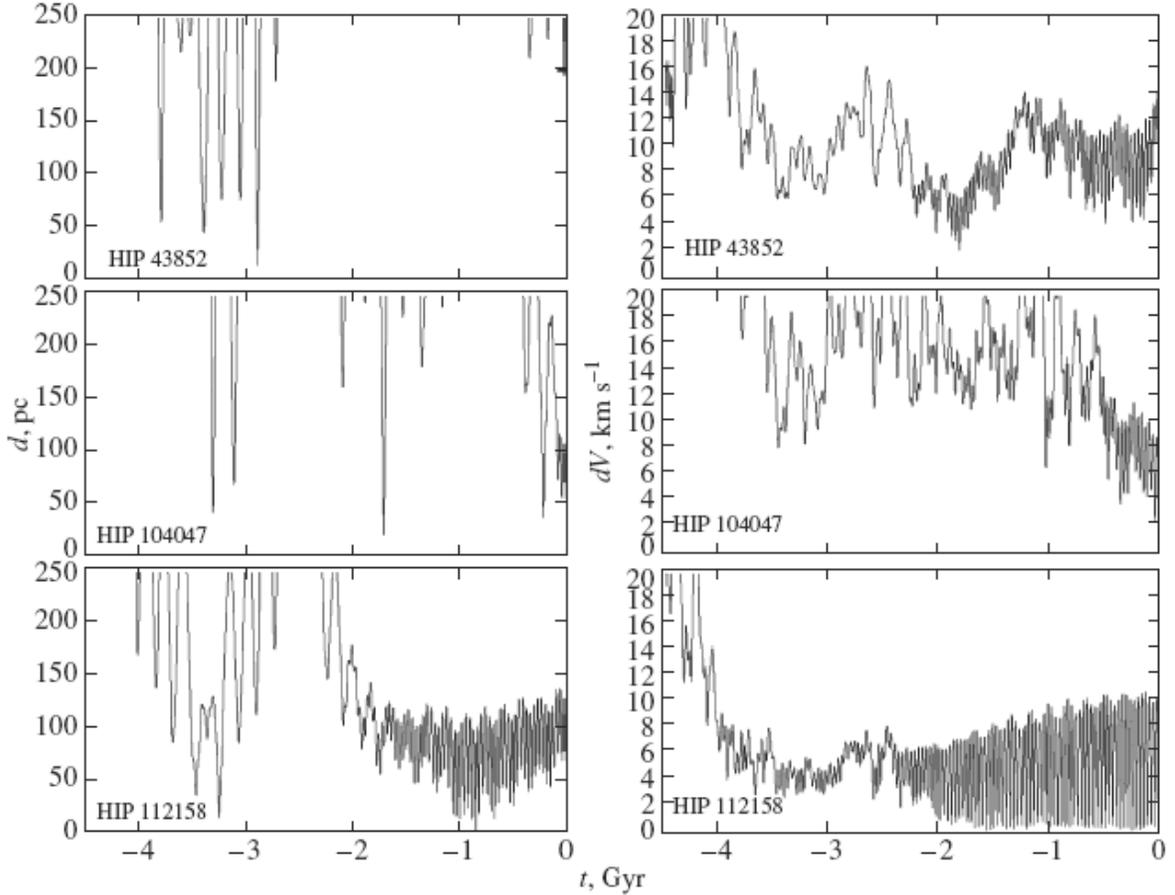}
 \caption{Parameters of the encounter $d$ and $dV$ of the three stars found with
the solar orbit obtained by taking into account the influence of
the (2 + 4) spiral pattern with the parameters (5) and (6) versus
time.
 }
\label{f3}
 \end{center}}
  \end{figure}

\subsection*{The Composite 2 + 4 Spiral Pattern Model}
The results of applying a more complex model of the spiral pattern
are of great interest. One of them is the composite (2 + 4) model
proposed by L\'{e}pine et al. (2001). In this model, the spiral
density wave potential (3) contains two terms with amplitudes A2
(two-armed component) and A4 (four-armed component). Here, the Sun
is on the corotation circle ($\Omega_0 = \Omega_p$) and, as was
shown by Mishurov and Acharova (2010), the influence of the spiral
pattern is so strong that the test model particles of the cluster
are scattered over a significant spatial volume.

The following parameters of the spiral density wave were adopted
for the two-armed spiral pattern:
\begin{equation}
 \begin{array}{lll}
 m=2,\\
 i=-7^\circ,\\
 f_{r0}=0.05,\\
 \chi_\odot=300^\circ,\\
 \Omega_p=27.5~\hbox {km s$^{-1}$ kpc$^{-1}$}
 \label{5}
 \end{array}
 \end{equation}
 and
 \begin{equation}
 \begin{array}{lll}
 m=4,\\
 i=-14^\circ,\\
 f_{r0}=0.035,\\
 \chi_\odot=140^\circ,\\
 \Omega_p=27.5~\hbox {km s$^{-1}$ kpc$^{-1}$}
 \label{6}
 \end{array}
 \end{equation}
for the four-armed spiral pattern. The results are presented in
Fig. 3. We can see that there are encounters of interest to us for
all three stars. However, these encounters are less close than
those obtained in the model of a four-armed spiral pattern with
parameters (4), whose results are presented in Fig. 2.

\section*{CONCLUSIONS}
From the XHIP catalogue (Anderson and Francis 2012), we selected
1872 F–G–K stars with relative parallax measurement errors
$\sigma_\pi/\pi<20\%$ and absolute values of their space
velocities relative to the Sun $<$15 km s$^{-1}$. We constructed
the Galactic orbits for all these stars for 4.5 Gyr into the past
using an axisymmetric Galactic potential model with allowance made
for the perturbations from the spiral density wave. The parameters
of the encounter with the solar orbit were calculated for each
orbit. For this purpose, we used the Galactic potential model
proposed by Fellhauer et al. (2006) for $R_0=8$ kpc and the model
of a four-armed spiral pattern.

We detected three new stars with Galactic orbits close to the
solar one in a long time interval into the past. These stars are
HIP 43852, HIP 104047, and HIP 112158. The high-mass spectroscopic
binary HIP 112158 is poorly suited for the role of the Sun’s
kinematic sibling by its age. This role is quite possible for the
single star HIP 43852 and the multiple system HIP 104047. We
confirm the status of our previously found candidates for close
encounters, HIP 47399 and HIP 87382.

For the three candidates revealed for the first time, HIP 43852,
HIP 104047, and HIP 112158, we modelled their encounters with the
Sun for the composite (2+4)model of the spiral pattern (L\'{e}pine
et al. 2001). This modeling showed the presence of encounters
approaches of interest to us, although they turned out to be less
close than those obtained in the model of a simple four-armed
spiral pattern with parameters (4).

In our opinion, the star HIP 87382 with a nearly solar chemical
composition is currently the most likely candidate, because it
persistently shows close encounters with the Sun on time scales of
more than 3 Gyr when using various axisymmetric Galactic potential
models both without and with allowance made for the influence of
the spiral density wave with parameters varying within their
measurement errors.

\section*{ACKNOWLEDGMENTS}
We are grateful to the referees for their useful remarks that
contributed to an improvement of the paper. This work was
supported by the “Origin and Evolution of Stars and Galaxies”
Program P-21 of the Presidium of the Russian Academy of Sciences
and the Program of State Support for Leading Scientific Schools of
the Russian Federation (project NSh-16245.2012.2, “Multiwavelength
Astrophysical Studies”). The work was performed using the SIMBAD
search database.

 \section*{REFERENCES}

{\small

 1. C. Allen and A. Santillan, Rev. Mex. Astron. Astrofis.
22, 255 (1991).

 2. E. Anderson and Ch. Francis, Astron. Lett. 38, 331
(2012).

3. S. F. A. Batista and J. Fernandes, New Astron. 17, 514 (2012).

4. J. Bland-Hawthorn and K. Freeman, Publ. Astron. Soc. Austral.
21, 110 (2004).

5. J. Bland-Hawthorn, M. R. Krumholz, and K. Freeman, Astrophys.
J. 713, 166 (2010).

6. V. Bobylev and A. Bajkova, Mon. Not. R. Astron. Soc. 408, 1788
(2010).

7. V. V. Bobylev, A. T. Bajkova, A. Myll\"{a}ri, and M. Valtonen,
Astron. Lett. 37, 550 (2011).

8. V. V. Bobylev and A. T. Bajkova, Astron. Lett. 38, 638 (2012).

9. V. V. Bobylev and A. T. Bajkova, Astron. Lett. 39, 809 (2013a).

10. V. V. Bobylev and A. T. Bajkova, Astron. Lett. 39, 759
(2013b).

11. V. V. Bobylev and A. T. Bajkova, Astron. Lett. 39, 532
(2013c).

12. V. Bobylev and A. Bajkova, Mon. Not. R. Astron. Soc. 437, 1549
(2014).

13. A. G. A. Brown, S. F. Portegies Zwart, and J. Bean, Mon. Not.
R. Astron. Soc. 407, 458 (2010).

14. L. Casagrande, R. Sch\"{o}nrich, M. Asplund, S. Cassisi, I.
Ramirez, J. Melendez, T. Bensby, and S. Feltzing, Astron.
Astrophys. 530, 138 (2011).

15. P. Demarque, J.-H. Woo, Y.-C. Kim, and K. Sukyoung, Astrophys.
J. Suppl. Ser. 155, 667 (2004).

16. M. Fellhauer, V. Belokurov, N. W. Evans, M. I. Wilkinson, D.
B. Zucker, G. Gilmore, M. J. Irwin, D. M. Bramich, et al.,
Astrophys. J. 651, 167 (2006).

17. D. Fernandez, F. Figueras, and J. Torra, Astron. Astrophys.
480, 735 (2008).

18. T. Foster and B. Cooper, ASP Conf. Ser. 438, 16 (2010).

19. O. Gerhard, Mem. Soc. Astron. Ital. Suppl. 18, 185 (2011).

20. G. A. Gontcharov, Astron. Lett. 32, 759 (2006).

21. L. Hernquist, Astrophys. J. 356, 359 (1990).

22. E. H\o g, C. Fabricius, V. V. Makarov, S. Urban, T. Corbin, G.
Wycoff, U. Bastian, P. Schwekendiek, and A.Wicenec, Astron.
Astrophys. 355, L27 (2000).

23. Y. C. Joshi, Mon. Not. R. Astron. Soc. 378, 768 (2007).

24. F. van Leeuwen, Astron. Astrophys. 474, 653 (2007).

25. J. R. D. L\'{e}pine, Yu. N. Mishurov, and S. Yu. Dedikov,
Astrophys. J. 546, 234 (2001).

26. C. C. Lin and F. H. Shu, Astrophys. J. 140, 646 (1964).

27. C. C. Lin, C. Yuan, and F. H. Shu, Astrophys. J. 155, 721
(1969).

28. O. Yu. Malkov, V. S. Tamazian, J. A. Docobo, and D. A.
Chulkov, Astron. Astrophys. 546, 69 (2012).

29. B. D. Mason, G. L. Wycoff, W. I. Hartkopf, G. G. Douglass, and
C. E. Worley, Astron. J. 122, 3466 (2001).

30. A. Massarotti, D. W. Latham, R. P. Stefanik, P. Robert, and J.
Fogel, Astron. J. 135, 209 (2008).

31. Yu. N. Mishurov and I. A. Acharova, Mon. Not. R. Astron. Soc.
412, 1771 (2011).

32. M. Miyamoto and R. Nagai, Publ. Astron. Soc. Jpn. 27, 533
(1975).

33. D. Montes, J. L\'{o}pez-Santiago, M. C. G\'{a}lvez, M. J.
Fernandez-Figueroa, E. De Castro, and M. Cornide, Mon. Not. R.
Astron. Soc. 328, 45 (2001).

34. T. Nakajima and J.-I. Morino, Astron. J. 143, 2 (2012).

35. B. Nordstr\"{o}m, M. Mayor, J. Andersen, J. Holmberg, F. Pont,
B. R. Jorgensen, E. H. Olsen, S. Udry, and N. Mowlavi, Astron.
Astrophys. 419, 989 (2004).

36. S. B. Parsons, Astron. J. 127, 2915 (2004).

37. S. Pfalzner, Astron. Astrophys. 549, 82 (2013).

38. M. E. Popova and A. V. Loktin, Astron. Lett. 31, 171 (2005).

39. S. F. Portegies Zwart, Astrophys. J. 696, L13 (2009).

40. D. Pourbaix, A. A. Tokovinin, A. H. Batten, F. C. Fekel, W. I.
Hartkopf, H. Levato, N. I. Morrell, G. Torres, and S. Udry,
Astron. Astrophys. 424, 727 (2004).

41. R. Sch\"{o}nrich, J. Binney, and W. Dehnen, Mon. Not. R.
Astron. Soc. 403, 1829 (2010).

42. A. Tokovinin, Mon. Not. R. Astron. Soc. 389, 925 (2008).

43. J. A. Valenti and D. A. Fischer, Astrophys. J. Suppl. Ser.
159, 141 (2005).

44. J. P. Vall\'{e}e, Astron. J. 135, 1301 (2008).

45. C. Yuan, Astrophys. J. 158, 889 (1969).

}

\end{document}